\documentclass[showkeys]{revtex4}
\include{epsf}

\begin{document}

\title{Novel Results on the Large-Signal Dynamic Admittance
of $p\!-\!n$-Junctions}
\author{  Edval J.\ P.\ Santos and Anatoly A.\ Barybin$^*$}
\affiliation{Laboratory for Devices and
Nanostructures at the Departamento de Eletr\^onica e Sistemas,
Universidade Federal de Pernambuco, 50740-530, Recife, Brazil.
E-mail: edval@ee.ufpe.br .\\
(*) Electronics Department,
Saint-Petersburg Electrotechnical University, 197376,
Saint-Petersburg, Russia, on leave at the Departamento de
Eletr\^onica e Sistemas, Universidade Federal de Pernambuco,
50740-530, Recife, Brazil. E-mail: barybin@mail.ru .}

\begin{abstract}
Recent theoretical results obtained by Barybin and Santos~\cite{1} have
suggested that the dynamic admittance of the $p\!-\!n$-junction is
proportional to the modified Bessel function of the first kind which
depend on an amplitude of ac signal. This result extends the
conventional theory usually encountered in known papers and textbooks.
In this letter, some experimental results are presented
to confirm our theoretical prediction. The measurements were performed
with a lock-in amplifier, using a low noise operational amplifier.
Two types of the $p\!-\!n$-diodes were employed to check our theory:
1N914B diode, typically used for high-frequency applications,
and 1N4007 diode, typically used in power supplies.
Experimental results are consistent with the theoretical ones
if a fitting parameter allowing for generation--recombination
processes in the depletion layer is taken into account.

\end{abstract}


\maketitle

\section{Introduction to theory}

Recently Barybin and Santos~\cite{1} have applied
a spectral approach to the theory of $p\!-\!n$-junctions in order
to take into account the effect of a large ac signal in a rigorous
theoretical manner.

This approach is based on the general
diffusion--drift equations for injected minority carriers, which
is a standard practice for semiconductor electronics.
The only specific feature distinguishing our approach from others
is related to the initial representation of desired carrier
concentrations in the form of Fourier expansion over frequency
harmonics. Such harmonics are produced by nonlinear processes in
$p\!-\!n$-junctions when a sufficiently large ac voltage amplitude
$V_1$ is applied to the junction together with the dc bias voltage
$V_0$. As a result, spectra of both the excess concentration of
injected carriers and the external circuit current have been
derived (see formulas (25), (26), and (32) of paper~\cite{1}).
The use of appropriate terms in Fourier series has given rise to
new expressions for the dc component $J_0$ and ac component $J_1$
of the external circuit current which determine the static
current--voltage characteristic $J_0(V_0)$ and the dynamic admittance
$Y=J_1/V_1$. They have both proved to be dependent not only on $V_0$
but also on $V_1$, in the case of nonlinear (in ac signal) regime of
operation of the $p\!-\!n$-junction.

Below we shall experimentally verify the novel theoretical result
concerning a dependence of the dynamic admittance
on the amplitude of applied ac voltage $V_\sim=2V_1$. This dependence
was missed by other authors beginning with
Shockley's original papers~\cite{2,3}. They all assume that the
amplitude of the ac voltage is much smaller than
$V_T$ ($V_\sim\ll V_T\equiv\kappa T/q$).  This approximation is also used
in popular books such as~\cite{4,5}.
The basic model of a diode admittance is a parallel combination
of the conductance $G_d$ and capacitance $C_d$. Based on the old theory,
the dynamic admittance of the $p\!-\!n$-diode is derived in the
form~\cite{4,5}
\begin{equation}
Y(\omega)= G_d^{old}(\omega) + i\omega C_d^{old}(\omega)
\label{eq:1}
\end{equation}
with the following {\it old\/} expressions for the diffusion
conductance and capacitance:
\begin{equation}
G_d^{old}(\omega)=
{ \sqrt{1+\sqrt{1+ \omega^2\tau_p^2}}\over\sqrt{2}}\;G_{d0} \,,
\label{eq:2}
\end{equation}
\begin{equation}
C_d^{old}(\omega)=
{ \sqrt{2}\over\sqrt{1+\sqrt{1+ \omega^2\tau_p^2}}}\;C_{d0} \,,
\label{eq:3}
\end{equation}
where $\tau_p$ is the minority carrier lifetime, and $\omega$ is
an operating frequency of the applied ac voltage $V_\sim$.
Expressions~(\ref{eq:2}) and~(\ref{eq:3}) are valid for the
$p^+\!-n$-structures with a highly doped $p$-emitter when
$p_n\gg\!n_p$ and the low-frequency diffusion conductance $G_{d0}$
and capacitance $C_{d0}$ are equal to~[5]
\begin{equation}
G_{d0}= \frac{qS}{\kappa T}\,\frac{qD_p p_n}{L_p}
\qquad\mbox{and}\qquad
C_{d0}= \frac{qS}{\kappa T}\,\frac{qL_p p_n}{2}\,,
\label{eq:4}
\end{equation}
where $L_p=\sqrt{D_p\tau_p}$ is the diffusion length for holes with
the equilibrium concentration $p_n$, and  $S$ is a cross-section area
of the diode.

Considering the low injection theory and applying the spectral
solution approach, we have found that for arbitrary applied voltage
levels expressions~(\ref{eq:2}) and~(\ref{eq:3}) should be modified
to the following {\it new\/} form (see Eqs.~(48) and (49)
of paper~[1]):
\begin{equation}
G_d^{new}(\omega,V_\sim)= G_d^{old}(\omega)\,
{I_1(\beta V_{\sim})\over \beta V_{\sim}/2} \,,
\label{eq:5}
\end{equation}
\begin{equation}
C_d^{new}(\omega,V_\sim)= C_d^{old}(\omega)\,
{I_1(\beta V_{\sim})\over \beta V_{\sim}/2} \,,
\label{eq:6}
\end{equation}
so that instead of the old expression (\ref{eq:1}) for the
dynamic admittance we have a new one:
\begin{equation}
Y(\omega,V_\sim)= G_d^{new}(\omega,V_\sim) +
i\omega C_d^{new}(\omega,V_\sim).
\label{eq:7}
\end{equation}

As follows from these formulas, the new diffusion conductance and
capacitance depend not only on the frequency~$\omega$ given by the
old formulas (\ref{eq:2}) and (\ref{eq:3}) but also on
the ac voltage amplitude $V_\sim$ appearing in the
modified Bessel function of first order $I_1(\beta V_{\sim})$,
where $\beta=q/\kappa T$. In the low-frequency limit when
$(\omega\tau_p)^2\!\ll 1$, equations (\ref{eq:2})--(\ref{eq:3}) and
(\ref{eq:5})--(\ref{eq:6}) reduce to the following normalized quantities:
\begin{equation}
{G_d^{new}(0,V_\sim)\over G_{d0}} =
{C_d^{new}(0,V_\sim)\over C_{d0}} =
{I_1(\beta V_{\sim})\over \beta V_{\sim}/2} \,.
\label{eq:8}
\end{equation}

At low signal amplitudes when $V_{\sim}\ll V_T$, the right-hand side
of Eq.~(\ref{eq:8}) is approximately equal to 1, which is the result
usually found in the literature (see, e.~g., formulas (64)--(66) and
Fig.~23 of book~[5]). Based on the new theory, even at zero bias it
is possible to get experimental evidence of the dynamic behavior
predicted by formula~(\ref{eq:8}).

Our objective is to investigate a dependence of the new diffusion
conductance on the ac voltage amplitude, $G_d^{new}(0,V_\sim)$,
at low frequencies and zero bias. Although the diffusion
capacitance $C_d^{new}(0,V_\sim)$ can also be measured,
we shall not considered it here because at zero bias the total
capacitance is dominated by the depletion capacitance. Below we present
the experimental results obtained from conductance measurements
to corroborate the new large-signal theory of $p\!-\!n$-junctions~[1].

\section{Experimental technique and results}

Our theory developed in paper~[1] is based on the ideal
Shockley model of the abrupt $p\!-\!n$-junction with neglecting
generation--recombination processes in the depletion layer~[3].
But in practice, such processes may be important for the actual
$p\!-\!n$-diodes employed below in our experiment. So they
have to be taken into account at least phenomenologically, as
suggested by Sze~[5].
Let us follow him and correct our theoretical results by an
empirical factor $n$ which is introduced so as to provide the
following replacement:
\begin{equation}
\beta\equiv\frac{q}{\kappa T} \to
\frac{q}{n\kappa T}\equiv\beta_n.
\label{eq:9}
\end{equation}

Values of the factor $n$ lie between 1 and 2: $n=1$ when a contribution
of the generation--recombination processes is negligibly small, and
$n=2$ if the recombination current dominates over the diffusion one~[5].

By using the replacement (\ref{eq:9}), expression (\ref{eq:8}) for the
low-frequency dynamic conductance can be rewritten in the following
corrected form:
\begin{equation}
G_d^{new}(0,V_\sim)= g_n(\beta V_{\sim})\,G_{d0},
\label{eq:10}
\end{equation}
where the correcting function $g_n(\beta V_{\sim})$ is defined as
\begin{equation}
g_n(\beta V_{\sim}) =
{I_1(\beta_n V_{\sim})\over \beta_n V_{\sim}/2}\equiv
{I_1(\beta V_{\sim}/n)\over \beta V_{\sim}/2n}\,.
\label{eq:11}
\end{equation}

In general, the modified quantity $\beta_n$ in expression (\ref{eq:11})
takes into account not only a contribution from the generation--recombination
processes by means of the factor $n$ but also that from a priori unknown
temperature $T$ of a $p\!-\!n$-junction under experimental investigation.
Hence, the product $nT$ can be used as a fitting parameter to adjust the
theoretical relations (\ref{eq:10})--(\ref{eq:11}) with experimental
results obtained below. The correcting function $g_n(\beta V_{\sim})$
expressed by formula~(\ref{eq:11}) is plotted in Fig.~1 for $T=300$\,K and
five values of the fitting parameter $n=1.0, 1.25, 1.5, 1.75$, and 2.0.

To verify our theory, we built a simple apparatus using a dual-phase DSP
lock-in amplifier, Stanford Research Systems model SR830, and a low
noise electrometer grade operational amplifier, Burr-Brown OPA\,128JM
(see Fig.~2). The operational amplifier is in the inverter configuration.
The ac voltage from the lock-in amplifier is directly applied to the
diode with no bias, and the output of the amplifier is connected
directly to the lock-in. The lock-in amplifier has two displays ---
X and Y which give the root-mean-square (rms) value of output signal
at the excitation frequency $\omega$. It is easy to show that the
lock-in output in the X and Y displays is equal to
\begin{equation}
V_{out}^X= -RV_{in} \bigl[
G_d(\omega)\cos(\phi-\phi_0) +
\omega C_d(\omega)\sin(\phi-\phi_0) \bigr],
\label{eq:12}
\end{equation}
\begin{equation}
V_{out}^Y= -RV_{in} \bigl[
G_d(\omega)\sin(\phi-\phi_0) +
\omega C_d(\omega)\cos(\phi-\phi_0) \bigr].
\label{eq:13}
\end{equation}
Here $R$ is a feedback resistor, $V_{in}$ is the rms value of
the lock-in oscillator voltage ($V_{in}=V_\sim/\sqrt{2}$),
$\phi$ is the internal phase of the lock-in oscillator signal
connected to the phase detector, and $\phi_0$
is the phase of the lock-in oscillator signal after passing through
an external circuit. From Eqs.~(\ref{eq:12}) and (\ref{eq:13}) it
follows that the two displays are $\pi/2$ out
of phase from each other.

To get correct experimental values, it is necessary to adjust a phase
$\phi$ of the lock-in local oscillator so that the X-display would be
used for the conductance voltage $|V_{out}^X|=RV_{in}G_d$ and the Y-display
for the capacitance voltage $|V_{out}^Y|=RV_{in}\omega C_d$. As follows
from expressions~(\ref{eq:12}) and (\ref{eq:13}), it can be realized
only if $\phi=\phi_0$. The phase $\phi$ is adjusted with a reference
capacitor $C$ which is placed between the local oscillator and the
minus input of the operational amplifier instead of diode~$D$,
as shown in Fig.~2. The phase adjustment is carried
out until zero voltage is observed in the conductance X-display,
i.~e., when $\phi=\phi_0$. Then Eqs.~(\ref{eq:12}) and (\ref{eq:13})
yield the required results:
\begin{equation}
G_d(\omega,V_\sim)= {1\over R}\,{|V_{out}^X|\over V_{in}}\,
\biggl|_{\,\phi\,=\,\phi_0}
\quad\mbox{and}\quad\;
\omega C_d(\omega,V_\sim)= {1\over R}\,{|V_{out}^Y|\over V_{in}}\,
\biggl|_{\,\phi\,=\,\phi_0}.
\label{eq:14}
\end{equation}

After the adjustment, the reference capacitor $C$ is replaced by a measured
diode $D$ and the measurements are performed by varying the applied ac
voltage. The operating frequency of the lock-in internal oscillator was
chosen 1\,kHz to surely provide the relation $\omega\tau_p\ll 1$
underlying the initial theoretical expression~(\ref{eq:8}).
To extract the conductance, one must normalize the data.

For our experiments we have employed two different diodes ---
the 1N914B, a common high frequency diode, and
the 1N4007, a diode used in power applications such as power
supplies.  All resistors and capacitors used in the experiment
were measured with the Stanford Research Systems model SR720 LCR
meter. The measurement error is found to be below $1\%$.

The experimental results for two above-mentioned diodes are presented
in Figs.~3 and~4.

As seen from Eqs.~(\ref{eq:10}) and (\ref{eq:11}), the low-voltage
measurements (when $g_n\!=1$) give values of the
low-frequency diffusion conductance $G_{d0}$ defined by formula
(\ref{eq:4}). These values have proved to be equal
$G_{d0}\simeq 6.4\times 10^{-7}$\,mho for the 1N914B diode and
$G_{d0}\simeq 3.7\times 10^{-7}$\,mho for the 1N4007 diode.

From the experimental curves plotted in Figs.~3 and~4 it follows
that the modified factor $\beta_n\equiv q/n\kappa T$ should be equal
to 20.6 for the 1N914B diode and 20.1 for the 1N4007 diode in order
to fit the theoretical expression~(\ref{eq:10}). Hence, the
fitting parameter $nT$ is respectively equal to 560\,K and
574\,K, which for the operating temperature $T=350$\,K
provides the following values of the generation--recombination
factor $n=1.6$ and $n=1.64$. Therefore, both the diodes operate in
a regime when the recombination current slightly dominates
over the diffusion current.

\section{Conclusion}

Application of a spectral approach to the charge carrier transport in
$p$-$n$-junctions with an arbitrary signal amplitude has allowed us to
obtain new results which were previously lost by all the authors.
Experimental results have corroborated theoretical predictions
of our large-signal theory.

The authors would like to thank the CNPq (Brazilian agency) for its
supporting this work. One of the authors (EJPS) would also like
to thank the ``Instituto do Mil\^{e}nio'' program funded by the CNPq.

\newpage
\begin{center}
\large{ Figure captions }
\end{center}

Fig.~1. The correcting function $g_n(\beta V_{\sim})$ defined by
formula~(\ref{eq:11}) and calculated for $T=300 K$ and five values
of the fitting parameter $n=1.0$ (curve~1), 1.25 (curve~2), 1.5 (curve~3),
1.75 (curve~4), and 2.0 (curve~5).

Fig.~2. Experimental setup with a measured diode $D$ whose dynamic
admittance $Y(\omega,V_\sim)$ is given by formula~(\ref{eq:7}).

Fig.~3. Theoretical curve and experimental data for the 1N914B diode.

Fig.~4. Theoretical curve and experimental data for the 1N4007 diode.

\newpage

\begin{figure}
        \epsfxsize=6in
        \centerline{\epsffile{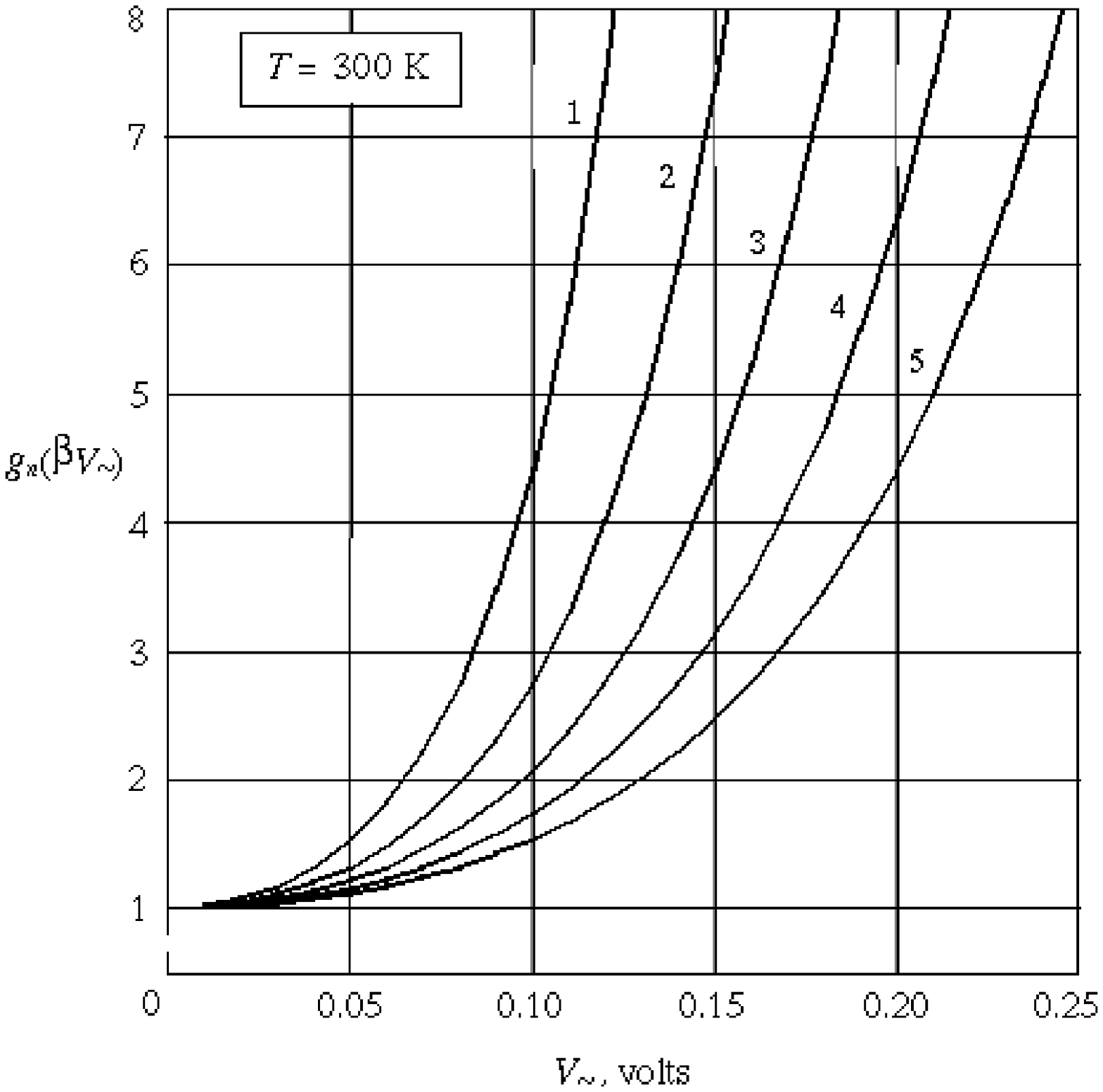}}
        \label{rayleigh}
        {\Large Fig .1 - Santos and Barybin}
\end{figure}

\newpage
 
\begin{figure}
        \epsfxsize=6in
        \centerline{\epsffile{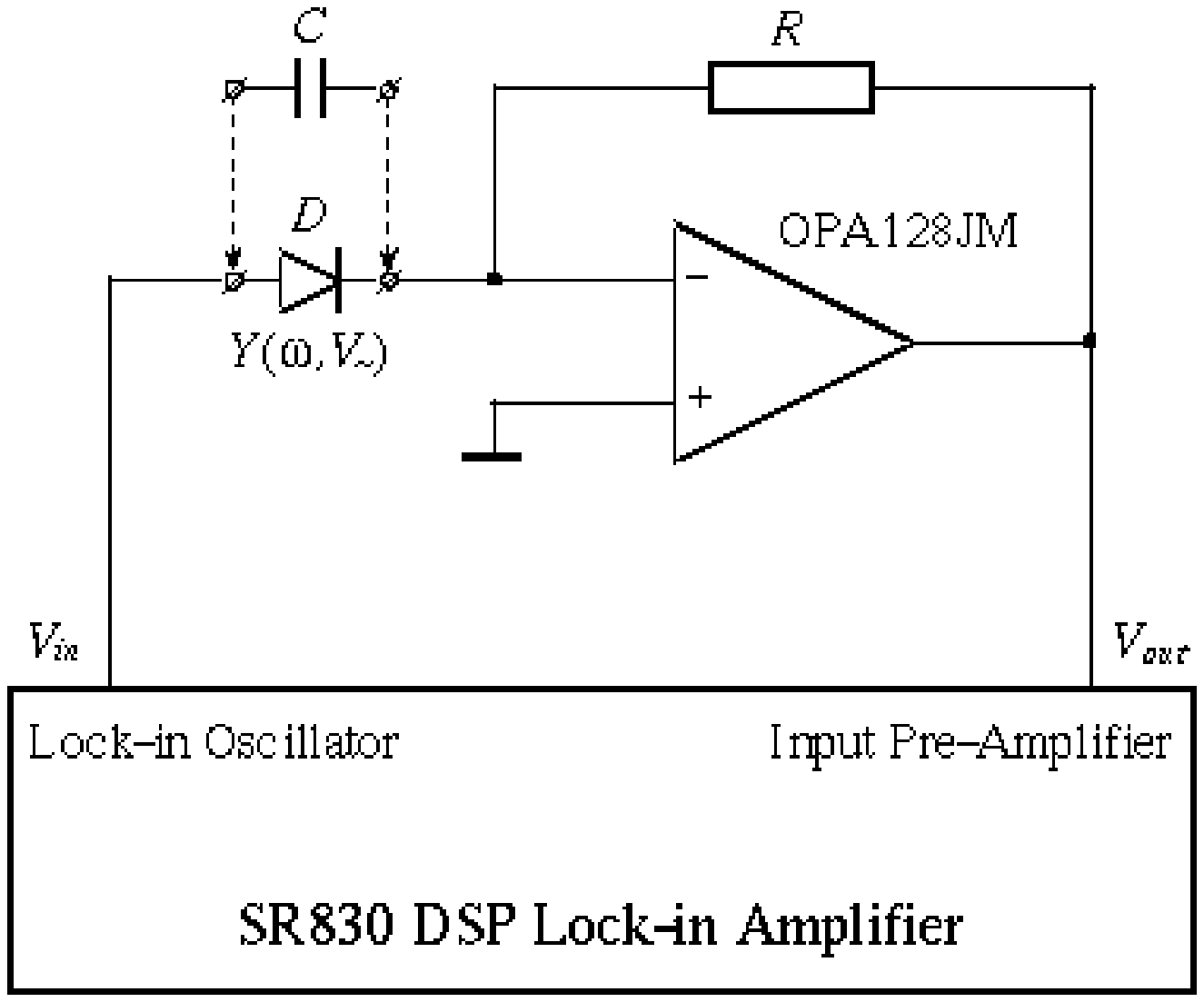}}
        {\Large Fig .2 - Santos and Barybin}
\end{figure}

\newpage
 
\begin{figure}
        \epsfxsize=6in
        \centerline{\epsffile{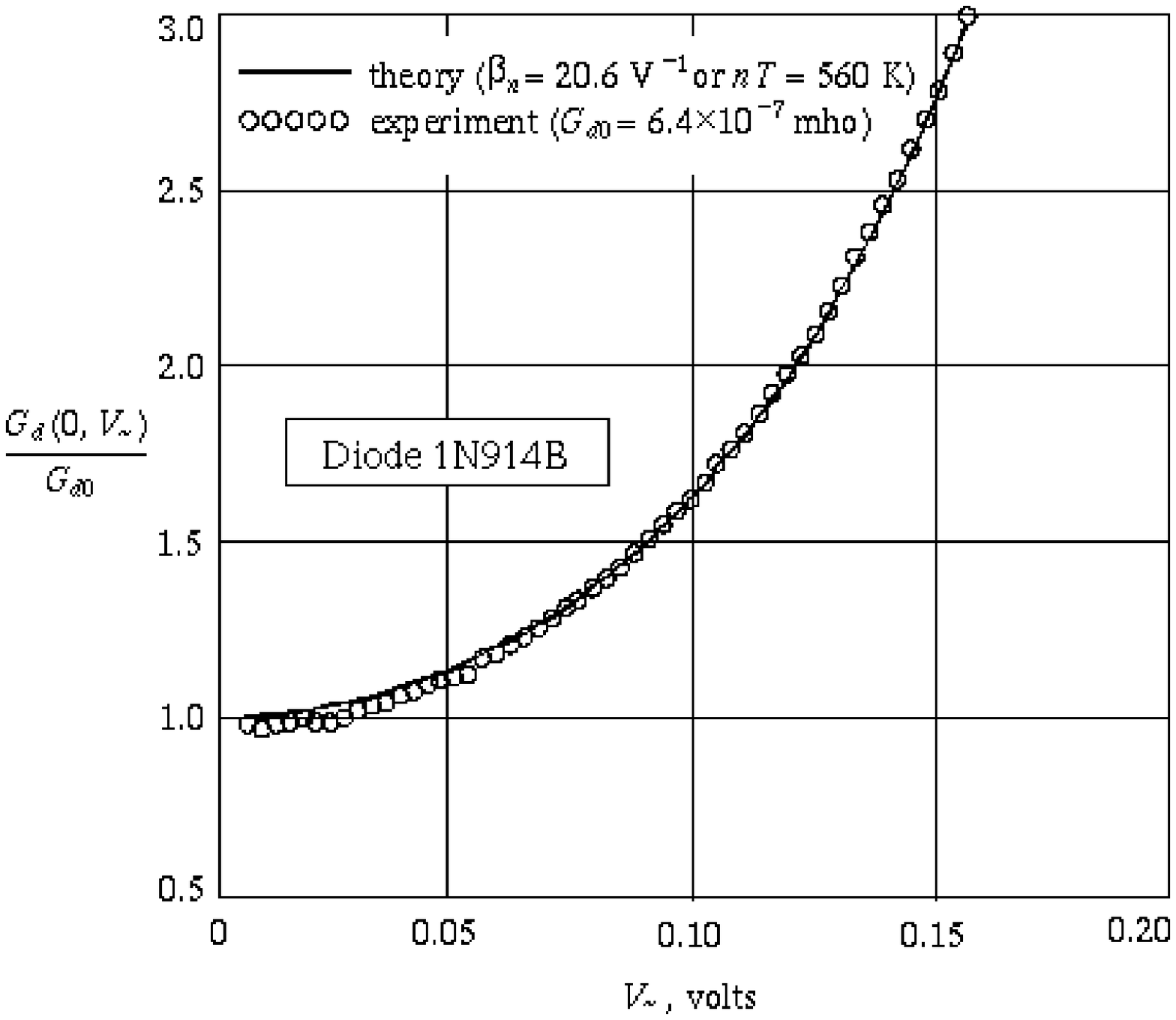}}
        {\Large Fig .3 - Santos and Barybin}
\end{figure}

\newpage
 
\begin{figure}
        \epsfxsize=6in
        \centerline{\epsffile{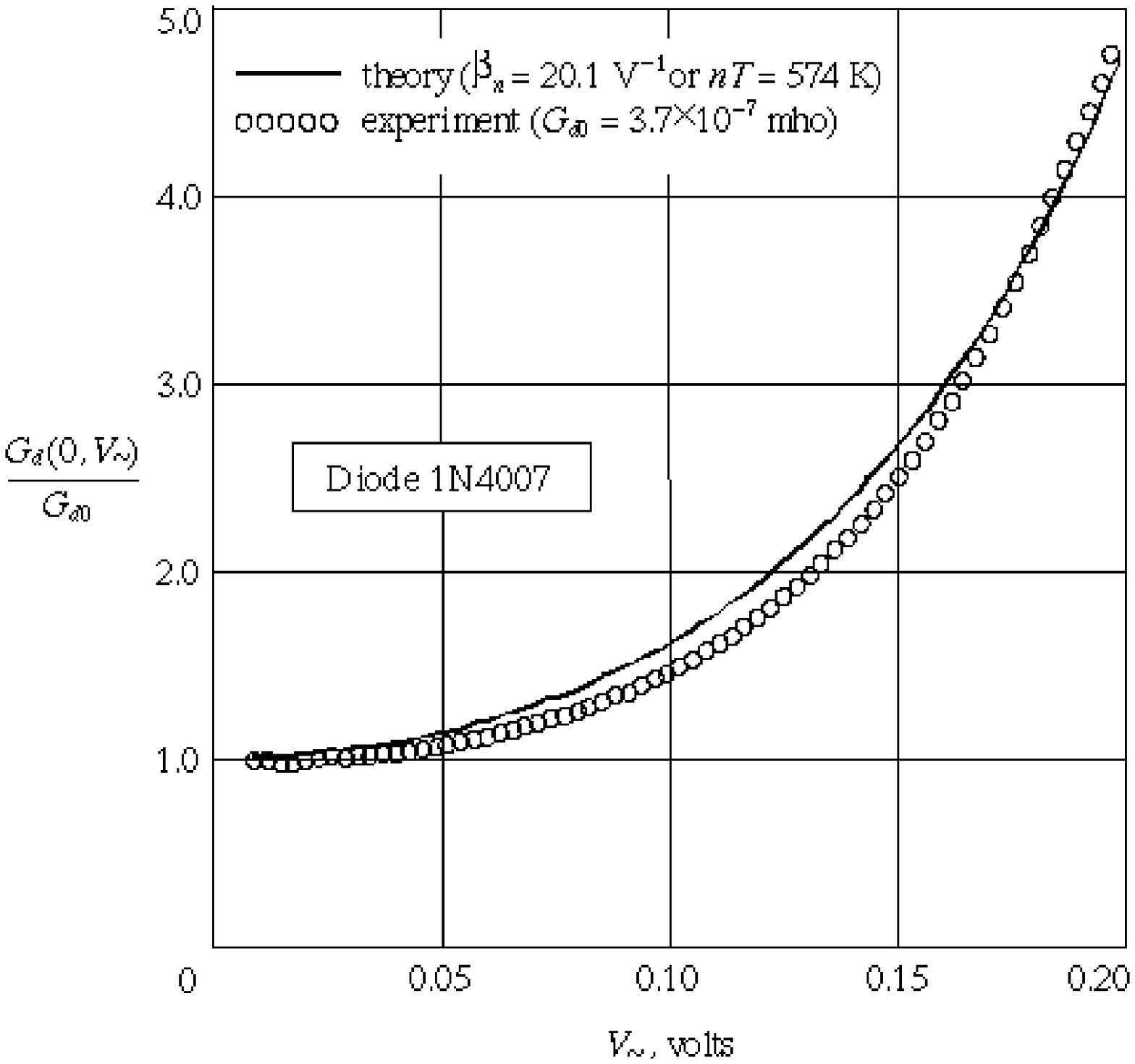}}
        {\Large Fig .4 - Santos and Barybin}
\end{figure}

\end{document}